\documentclass[12pt,english,floatfix,superscriptaddress,aps,prd,preprint]{revtex4}

\usepackage{graphicx}
\usepackage{dcolumn}
\usepackage{bm}
\usepackage[utf8]{inputenc}
\usepackage{slashed}
\usepackage{hyperref}
\usepackage[utf8]{inputenc}
\usepackage{float}
\usepackage{array}
\usepackage{lipsum}
\usepackage{graphicx}
\usepackage{amsmath,amsthm,amsfonts,amssymb}
\usepackage{graphicx}
\usepackage[english]{babel}
\usepackage{color}
\usepackage{esint}
\usepackage[dvips]{epsfig}
\usepackage[dvips]{graphicx}
\usepackage{float}
\usepackage{units}
\usepackage{textcomp}
\usepackage{mathrsfs}
\usepackage{amsmath}
\usepackage[makeroom]{cancel}
\usepackage{amssymb}
\usepackage{amsbsy}
\usepackage{amsfonts}

\usepackage{hyperref}
\hypersetup{colorlinks,breaklinks,
			citecolor=[rgb]{0.0,0.0,1.0},
            urlcolor=[rgb]{0.0,0.0,1.0},
            linkcolor=[rgb]{0,0.7,0.9}}
\usepackage{slashed}
\usepackage{slashed}

\newcommand{\ie}{\begin{equation}}
\newcommand{\fe}{\end{equation}}
\newcommand{\se}{\begin{eqnarray}}
\newcommand{\ff}{\end{eqnarray}}

\begin{document}

\title{Dual equivalence between self-dual and topologically massive $B\wedge F$ models coupled  to matter in $3+1$ dimensions}

\author{R. V. Maluf}
\email{r.v.maluf@fisica.ufc.br}
\affiliation{Universidade Federal do Cear\'a (UFC), Departamento de F\'isica,\\ Campus do Pici, Fortaleza,  CE, C.P. 6030, 60455-760 - Brazil.}

\author{F. A. G. Silveira}
\email{adevaldo.goncalves@fisica.ufc.br}
\affiliation{Universidade Federal do Cear\'a (UFC), Departamento de F\'isica,\\ Campus do Pici, Fortaleza,  CE, C.P. 6030, 60455-760 - Brazil.}

\author{J. E. G. Silva}
\email{euclides.silva@ufca.edu.br}
\affiliation{Universidade Federal do Cariri (UFCA), Av. Tenente Raimundo Rocha, Cidade Universit\'{a}ria, Juazeiro do Norte, Cear\'{a}, CEP 63048-080, Brazil.}

\author{C. A. S. Almeida}
\email{carlos@fisica.ufc.br}
\affiliation{Universidade Federal do Cear\'a (UFC), Departamento de F\'isica,\\ Campus do Pici, Fortaleza,  CE, C.P. 6030, 60455-760 - Brazil.}


\date{\today}


\begin{abstract}
In this work, we revisit the duality between a self-dual non-gauge invariant theory and a topological massive theory in $3+1$ dimensions. The self-dual Lagrangian is composed by a vector field and an antisymmetric field tensor whereas the topological massive Lagrangian is build using a $B \wedge F$ term. Though the Lagrangians are quite different, they yield to equations of motion that are connected by a simple dual mapping among the fields. We discuss this duality by analyzing the degrees of freedom in both theories and comparing their propagating modes at the classical level. Moreover, we employ the master action method to obtain a fundamental Lagrangian that interpolates between these two theories and makes evident the role of the topological $B \wedge F$ term in the duality relation. By coupling these theories with matter fields, we show that the duality holds provided a Thirring-like term is included. In addition, we use the master action in order to probe the duality upon the quantized fields. We carried out a functional integration of the fields and compared the resulting effective Lagrangians.
\end{abstract}

\maketitle

\section{Introduction}

Dualities are a main theme in nowadays physics. By connecting different theories or opposite regimes of a same model, dualities are powerful tools to seek and understand new effects. Notably, string theories are connected by $T$ and $S$ dualities \cite{tduality,Polchinski} and the $AdS/CFT$ correspondence links low-energy gravitational theory in $AdS$ spacetime with a strong coupling regime of a conformal field theory at the boundary \cite{maldacena}. Among the duality processes, the so-called bosonisation is of special importance and widely used to investigate nonperturbative properties in quantum field theory and condensed matter systems in low dimensions \citep{BurgessQuevedo}. In $1+1$ dimension, it is possible to establish a fermion-boson correspondence based on the properties of the Fermi surfaces \cite{Mandelstam}. This duality can be further generalized for non-abelian fields \cite{Witten84} and even for higher dimensions \cite{Marino, Burgess}. Recently, the bosonization lead to new $2+1$ relations called web of dualities \cite{Hernaski1,Hernaski2}.

Another example of duality involves topologically massive gauge theories. A well-known duality occurs between the self-dual (SD) \cite{auto dual original} and the Maxwell-Chern-Simons (MCS) \cite{MCS auto dual} models. These two theories describe a single massive particle of spin-1 in $2+1$ dimensional Minkowski space-time. Nevertheless, only the MCS model is gauge-invariant. The equivalence between the SD and MCS models was initially proved by Deser and Jackiw \cite{MCS auto dual}, and over the years, several studies of this equivalence have been carried out in the literature \cite{Karlhede1,Fradkin,Bralic,Banerjee,Banerjee2,malacarne,Minces,Anacleto2001}.
Particularly, by considering couplings with fermionic fields, it was shown in \cite{malacarne} that the models are equivalent provided that a Thirring-like interaction is included. In addition, supersymmetric \cite{Karlhede2,Ferrari1,Ferrari2} and noncommutative \cite{Gomes} extensions to the duality involving the SD and MCS models have been studied in different contexts.

At the heart of this duality, the Chern-Simons term plays a key role. An alternative topological term in $3+1$ dimensions can be formed from a $U(1)$ vector gauge field $A_{\mu}$ and a rank-2 antisymmetric tensor field $B_{\mu\nu}$, also known as the Kalb-Ramond field \cite{KalbRamon,Green}. Such a massive topological term is commonly called the $B \wedge F$ term \cite{BFOriginal,MKR,Oda,Lahiri}. Therefore, a natural generalization of the MCS model in four dimensions consists of the Maxwell and Kalb-Ramon fields coupled by a $B \wedge F$ term \cite{Taegyu}. This topologically massive gauge-invariant $B \wedge F$ theory ($TM_{B \wedge F}$) is unitary and renormalizable when minimally coupled to fermions, and represents a massive particle of spin-1 \cite{BFOriginal}. Models involving the Kalb-Ramond field have been extensively studied in the literature, specially in connection with string theories \cite{Soo}, quantum field theory \cite{Deguchi1999, Hari}, supersymmetry \cite{Almeida},  Lorentz symmetry violation \cite{Altschul2010,Hernaski2016,Maluf2018,Mariz2019}, black hole solutions \cite{Euclides2020}, cosmology \cite{Grezia}, and brane words scenarios \cite{Wilami1,Wilami2}.

A self-dual version of the $TM_{B \wedge F}$ model was studied in Ref. \cite{BF_dual}. It involves the $B \wedge F$ term in a non-gauge invariant, first-order model ($SD_{B \wedge F}$). Such work showed the classic equivalence between the models, i.e., at the level of the equations of motion, through the gauge embedding procedure \cite{Anacleto2001}. In addition, when interactions with fermionic fields are considered, the duality mapping only is preserved if Thirring-like terms are taken into account, analogously to the SD/MCS case in $2+1$ dimensions. Yet, the issues regarding the generalization for arbitrary non-conserved matter currents and the proof of quantum duality have not yet been fully elucidated.
  
The main goal of this work is to provide an alternative method, via master action \cite{Dalmazi2009}, to prove the duality between the $SD_{B \wedge F}$ and $TM_{B \wedge F}$ theories, when the fields of the SD sector couple linearly with non-conserved currents, composed by arbitrary dynamic fields of matter. The master action approach has the advantage of providing a fundamental theory that interpolates between the two models and allows a more direct demonstration of duality at the quantum level. Besides, the master action method is a natural trail for the supersymmetric generalization of the duality studied here \cite{Ferrari2}.

The present work is organized as follows. In section \ref{section2}, we present the $SD_{B \wedge F}$ and $TM_{B \wedge F}$ theories in the free case, review their main physical characteristics, and check the classic duality by comparing their equations of motion. Moreover, we built a master Lagrangian density from the $TM_{B \wedge F}$ model, introducing auxiliary fields in order to obtain a first-order derivative theory. In section \ref{section3}, we included matter couplings in the SD sector and verify whether the equivalence is still compatible. We apply our results to the case of minimal coupling with fermionic matter and compare it with those found in the literature. In section \ref{section4}, we investigate the equivalence at the quantum level within the path-integral framework. Finally in section \ref{conclusion} we provide our conclusions and perspectives concerning further investigations.

\section{The duality at the classical level.}\label{section2}

In a $2+1$ flat spacetime Townsend, Pilch, and Nieuwenhuizen proposed a first-order derivative theory self-dual to the topological Chern-Simons theory \cite{auto dual original}. In four dimensions, this kind of duality can be built through a topological $B \wedge F$  term. In fact, consider a gauge non-invariant $SD_{B \wedge F}$ model composed by a vector field $A_{\mu}$ and an antisymmetric 2-tensor field $B_{\mu\nu}$ governed by the Lagrangian density \cite{BFOriginal, BF_dual}
\begin{align}\label{SDmodel}
 \mathcal{L}_{SD}
 =
 \frac{m^{2}}{2}A_{\mu}A^{\mu} - \frac{1}{4}B_{\mu\nu}B^{\mu\nu} + \frac{\chi\theta}{4}\epsilon_{\mu\nu\alpha\beta}B^{\mu\nu}F^{\alpha\beta},
 \end{align}
where $m$ is a parameter with dimension of mass, $\theta$ is a dimensionless coupling constant and $\chi=\pm 1$ defines either the self-duality $(+)$ or the anti self-duality $(-)$ to the theory. The field strengths associated with the vector and tensor fields are defined respectively by $F_{\mu\nu} = \partial_{\mu} A_{\nu} - \partial_{\nu}A_{\mu} $ and $H_{\mu\nu\alpha}= \partial_{\mu} B_{\nu\alpha} + \partial_{\nu} B_{\alpha\mu} + \partial_{\alpha} B_{\mu\nu}$. The equations of motion for the $A_{\mu}$ and $B_{\mu\nu}$ fields are, respectively,
\begin{eqnarray}
m^{2}A_{\beta}-\frac{\chi\theta}{2}\epsilon_{\mu\nu\alpha\beta}\partial^{\alpha}B^{\mu\nu}&=&0,\label{eq1}\\
B_{\mu\nu}-\chi\theta\epsilon_{\mu\nu\alpha\beta}\partial^{\alpha}A^{\beta}&=&0,\label{eq2}
\end{eqnarray}
and satisfy the constraint relations
\begin{eqnarray}
\partial_{\mu}A^{\mu}&=&0,\label{constraint1}\\
\partial^{\mu}B_{\mu\nu}&=&0.\label{constraint2}
\end{eqnarray}
Eqs. (\ref{eq1}) and (\ref{eq2}) form a set of coupled first-order differential equations that can be rewritten, with the help of relations (\ref{constraint1}) and (\ref{constraint2}), in the form of a wave equation given by
\begin{equation}
\left[\Box+\frac{m^{2}}{\theta^{2}}\right]\varphi=0,
\end{equation} where $\varphi$ denotes $A_{\mu}$ or $B_{\mu\nu}$ fields. This implies that the first-order Lagrangian density $\mathcal{L}_{SD}$  describes the dynamics of a massive vector field. In fact, the field $B_{\mu\nu}$ is auxiliary and can be removed from the action leading to \cite{Altschul2010} 
\begin{equation}
\mathcal{L}_{SD}
 = \frac{m^{2}}{2}A_{\mu}A^{\mu} - \frac{\theta^{2}}{4}F_{\mu\nu}F^{\mu\nu},
\end{equation}which is the Lagrangian density for a massive vector field with three propagating degrees of
freedom.

In the context of the present work, we are interested in investigate the equivalence between the self-dual model \eqref{SDmodel} and a second-order gauge-invariant theory. For this purposes, let us consider a topologically massive  $B \wedge F$ model defined as  \cite{MKR,BF_dual} 
\begin{align}
\mathcal{L}_{TM}
=
\frac{\theta^{2}}{12m^{2}}H_{\mu\nu\alpha}H^{\mu\nu\alpha} - \frac{\theta^{2}}{4}F_{\mu\nu}F^{\mu\nu} - \frac{\chi\theta}{4}\epsilon_{\mu\nu\alpha\beta}B^{\mu\nu}F^{\alpha\beta}.\label{mkr}
\end{align}
Note that the first two terms of $\mathcal{L}_{TM}$ are invariant under the gauge transformations $A_{\mu}\rightarrow A_{\mu}+\partial_{\mu}\lambda$ and $B_{\mu\nu}\rightarrow B_{\mu\nu}+\partial_{\mu}\beta_{\nu}-\partial_{\nu}\beta_{\mu}$, whereas the variation of the last term yields to a total divergence. The gauge parameter $\beta_{\mu}$ still has a subsidiary gauge transformation $\beta_{\mu}\rightarrow \beta_{\mu}+\partial_{\mu}\alpha$ that leaves $B_{\mu\nu}$ unchanged. The equations of motions derived from this Lagrangian density are
\begin{eqnarray}
\frac{\theta^{2}}{2m^{2}}\partial^{\mu}H_{\mu\nu\lambda}+\frac{\chi\theta}{4}\epsilon_{\nu\lambda\alpha\beta}F^{\alpha\beta}&=&0,\label{em-mkr1}\\
\theta^{2}\partial^{\mu}F_{\mu\lambda}+\frac{\chi\theta}{6}\epsilon_{\mu\nu\alpha\lambda}H^{\alpha\mu\nu}\label{em-mkr2}
&=&0.
\end{eqnarray}

In general, the two fields $A_{\mu}$ and $B_{\mu\nu}$ have four
and six independent degrees of freedom, respectively. However, due to the gauge symmetry in the theory described by $\mathcal{L}_{TM}$, some of them can be eliminated. In order to identify which ones propagate as massive physical modes
or which are spurious (gauge dependent) modes, it is instructive to perform
a decomposition in time-space on the equations of motions (\ref{em-mkr1}) and
(\ref{em-mkr2}). For this purpose, let us split $B_{\mu\nu}$ into the independent
components $B_{0i}$ and $B_{ij}$ and to introduce spatial vectors
$\vec{\mathcal{X}}$ and $\vec{\mathcal{Y}}$ defined by 
\begin{equation}
\mathcal{X}^{i}\equiv-B_{0i},\ \ \ \ \ \mathcal{Y}^{i}\equiv\frac{1}{2}\epsilon^{ijk}B_{jk},
\end{equation}
where $\epsilon^{0ijk}=\epsilon^{ijk}$. With these definitions, we obtain a set of coupled second order differential equations in the form
\begin{eqnarray}
\nabla^{2}A^{0}+\partial^{0}\partial_{i}A^{i}+\frac{\chi}{\theta}\partial_{i}\mathcal{Y}^{i}	&=&0,\\
\Box A^{i}-\partial^{i}\left(\partial_{0}A^{0}+\partial_{j}A^{j}\right)+\frac{\chi}{\theta}\left(\epsilon^{ijk}\partial_{k}\mathcal{X}_{j}+\partial_{0}\mathcal{Y}^{i}\right)	&=&0,\\
-\nabla^{2}\mathcal{X}_{i}-\partial_{i}\partial^{j}\mathcal{X}_{j}+\epsilon_{ijk}\left(\partial_{0}\partial^{j}\mathcal{Y}^{k}-\frac{\chi m^{2}}{\theta}\partial^{j}A^{k}\right)	&=&0,\\
\partial_{0}^{2}\mathcal{Y}^{k}+\partial_{i}\partial^{k}\mathcal{Y}^{i}+\epsilon^{ijk}\partial_{0}\partial_{j}\mathcal{X}_{i}+\frac{\chi m^{2}}{\theta}\left(\partial^{k}A^{0}-\partial^{0}A^{k}\right)	&=&0.
\end{eqnarray}

After some manipulation of these equations, we can formally solve the temporal component $A^{0}$
and the 3-vector $\vec{\mathcal{X}}$ in terms of the other components according to 
\begin{eqnarray}
A^{0}&=&-\frac{1}{\nabla^{2}}\left(\partial^{0}\partial_{i}A_{(L)}^{i}+\frac{\chi}{\theta}\partial_{i}\mathcal{Y}_{(L)}^{i}\right),\\
\mathcal{X}_{i}^{(T)}&=&\frac{1}{\nabla^{2}}\epsilon_{ijk}\left(\partial_{0}\partial^{j}\mathcal{Y}_{(T)}^{k}-\frac{\chi m^{2}}{\theta}\partial^{j}A_{(T)}^{k}\right),
\end{eqnarray}where $v_{(T)}^{i}\equiv\theta_{j}^{i}v^{j}$ and $v_{(L)}^{i}\equiv\omega_{j}^{i}v^{j}$
are the transversal $(T)$ and longitudinal $(L)$ components of a
3-vector $\vec{v}$, respectively, with the projectors $\theta_{j}^{i}$
and $\omega_{j}^{i}$ defined by
\begin{equation}
\theta_{j}^{i}\equiv\delta_{j}^{i}-\omega_{j}^{i},\ \ \ \ \ \omega_{j}^{i}\equiv-\frac{\partial_{j}\partial^{i}}{\nabla^{2}}.
\end{equation}
Similar procedures can be applied to the components of the $\vec{A}$ and $\vec{\mathcal{Y}}$, such that
\begin{eqnarray}
\left[\Box+\frac{m^{2}}{\theta^{2}}\right]A_{(T)}^{i}&=0,\\
\left[\Box+\frac{m^{2}}{\theta^{2}}\right]\mathcal{Y}_{(L)}^{i}&=0.
\end{eqnarray}
The form of these solutions reveals that the only physical components are $A^{i}_{(T)}$ and $\mathcal{Y}_{(L)}^{i}$, while the other are auxiliary or gauge modes. Furthermore, as the longitudinal part of $\vec{\mathcal{Y}}$ is curl-free, it propagates as a massive scalar field, i.e., $\vec{\mathcal{Y}}=\nabla\phi$, whose mass depends on the coupling constant $\theta$. Thus, the results above show that the $TM_{B\wedge F}$ theory defined in (\ref{mkr}), like the  $SD_{B\wedge F}$ model, contains three massive propagating modes.

To make explicit the hidden duality between the models described above, it is convenient to introduce the dual fields associated with the field strength tensors $H^{\mu\nu\alpha}$ and $F^{\mu\nu}$, respectively by 
\begin{eqnarray}
\tilde{H}_{\mu}&\equiv & -\frac{\chi\theta}{6m^{2}}\epsilon_{\mu\nu\alpha\beta}H^{\nu\alpha\beta},\label{DualH}\\
\tilde{F}_{\mu\nu} &\equiv &\frac{\chi\theta}{2}\epsilon_{\mu\nu\alpha\beta}F^{\alpha\beta}\label{DualF}.
\end{eqnarray}
In terms of $\tilde{H}_{\mu}$ and $\tilde{F}_{\mu\nu}$,  the equations of motion (\ref{em-mkr1}) and (\ref{em-mkr2}) become
\begin{eqnarray}
m^{2}\tilde{H}_{\beta}
-
\frac{\theta}{2\chi}\epsilon_{\mu\nu\alpha\beta}\partial^{\mu}\tilde{F}^{\nu\alpha}
&=&0,\label{em-mkr1.1}
\\
\tilde{F}_{\mu\nu}
-
\frac{\theta}{\chi}\epsilon_{\mu\nu\alpha\beta}\partial^{\alpha}\tilde{H}^{\beta}
&=&0.\label{em-mkr2.1}
\end{eqnarray}
A direct comparison between the pairs of equations (\ref{eq1},\ref{eq2}) and (\ref{em-mkr1.1},\ref{em-mkr2.1}) shows that the dual fields $\tilde{H}_{\beta}$ and $\tilde{F}_{\mu\nu}$ satisfy exactly the same equations obtained for $SD_{B \wedge F}$ model when we identify $A_{\mu}\rightarrow \tilde{H}_{\mu}$ and $B_{\mu\nu}\rightarrow\tilde{F}_{\mu\nu}$. Therefore, the basic fields of the $SD_{B \wedge F}$ model correspond to the dual fields of the $TM_{B \wedge F}$ model. This proves the classical equivalence via equations of motion in the free field case. 

However, despite having established the dual connection, the mapping $A_{\mu}\rightarrow \tilde{H}_{\mu}$ and $B_{\mu\nu}\rightarrow\tilde{F}_{\mu\nu}$ leads to
\begin{equation}
\mathcal{L}_{TM}(\tilde{H},\tilde{F})=-\frac{m^{2}}{2}\tilde{H}_{\mu}\tilde{H}^{\mu}+\frac{1}{4}\tilde{F}_{\mu\nu}\tilde{F}^{\mu\nu}-\frac{1}{2}B_{\mu\nu}\tilde{F}^{\mu\nu},\label{Lmkr2}
\end{equation}wherein the identities $ F_{\mu\nu}F^{\mu\nu}=-1/\theta^{2}\tilde{F}_{\mu\nu}\tilde{F}^{\mu\nu}$ and $H_{\mu\nu\alpha}H^{\mu\nu\alpha}=-6m^{4}/\theta^{2}\tilde{H}_{\mu}\tilde{H}^{\mu}$ were used. 
Note that (\ref{Lmkr2}) does not recover (\ref{SDmodel}) and the equivalence between the two models is not evident. 
The common origin of these Lagrangian densities can be better addressed by means of the \textit{master Lagrangian} method, which we will formulate in the sequel.

\subsection{Classic Duality via Master Lagrangian}

The study of dual equivalence among four-dimensional models containing a topological $B\wedge F$ term was carried out for the first time in Ref. \cite{BF_dual}, whereby the authors used the dynamical gauge embedding formalism to show the classic duality between (\ref{SDmodel}) and (\ref{mkr}). Here, we employ the master Lagrangian method \cite{MCS auto dual, malacarne} that extends and interpolates those two studied models. Moreover, this method allows us to study the duality at the quantum level more directly.

Let us start from Lagrangian density $\mathcal{L}_{TM}$ in the form (\ref{Lmkr2}) written explicitly in terms of the fundamental fields $A_{\mu}$ and $B_{\mu\nu}$. Following \cite{MCS auto dual}, we will introduce auxiliary fields $\Pi_{\mu}$ and $\Lambda_{\mu\nu}$ in order to obtain a first-order derivative theory such that
\begin{equation}
\mathcal{L}_{M}=a\Pi_{\mu}\epsilon^{\mu\rho\sigma\delta}\partial_{\rho}B_{\sigma\delta}+b\Pi_{\mu}\Pi^{\mu}+c\Lambda_{\mu\nu}\epsilon^{\mu\nu\rho\sigma}\partial_{\rho}A_{\sigma}+d\Lambda_{\mu\nu}\Lambda^{\mu\nu}-\frac{\chi\theta}{2}\varepsilon_{\mu\nu\alpha\beta}B^{\mu\nu}\partial^{\alpha}A^{\beta},\label{Lmaster}
\end{equation}where $a$, $b$, $c$ and $d$ are constant coefficients to be determined. Note that the presence of mass terms for $\Pi_{\mu}$ and $\Lambda_{\mu\nu}$ ensures the auxiliary character of these fields.

The functional variation of $\mathcal{L}_{M}$ with respect to the auxiliary fields $\Pi_{\mu}$ and $\Lambda_{\mu\nu}$ allows us to write
\begin{eqnarray}
\Pi_{\mu} &=& -\frac{a}{2b}\epsilon_{\mu\nu\alpha\beta}\partial^{\nu}B^{\alpha\beta},\label{Pi}\\
\Lambda_{\mu\nu}	&=&-\frac{c}{2d}\epsilon_{\mu\nu\alpha\beta}\partial^{\alpha}A^{\beta}.\label{Gamma}
\end{eqnarray}

Substituting (\ref{Pi}) and (\ref{Gamma}) in (\ref{Lmaster}) and imposing $\mathcal{L}_{M}=\mathcal{L}_{TM}$, we obtain the relations
\begin{eqnarray}
\frac{a^{2}}{b}	&=&\frac{\theta^{2}}{2m^{2}},\\
\frac{c^{2}}{d}	&=&-\theta^{2}.
\end{eqnarray}

The same procedure can be performed for the fields $A_{\mu}$ and $B_{\mu\nu}$, and we can immediately solve their equations of motion, obtaining the following solutions:
\begin{eqnarray}
A_{\mu}	&=&\frac{2a}{\chi\theta}\Pi_{\mu}+\partial_{\mu}\phi,\label{A}\\
B_{\mu\nu}&=& \frac{2c}{\chi\theta}\Lambda_{\mu\nu}+\partial_{\mu}\Sigma_{\nu}-\partial_{\nu}\Sigma_{\mu}\label{B},
\end{eqnarray}being $\phi$ and $\Sigma_{\mu}$ arbitrary fields. Now, replacing (\ref{A}) and (\ref{B}) in (\ref{Lmaster}) and imposing $\mathcal{L}_{M}= \mathcal{L}_{SD}$, we obtain
\begin{eqnarray}
b &=&\frac{m^{2}}{2},\\
d &=&-\frac{1}{4},
\end{eqnarray}such that we can immediately fix $a=c=\chi\theta/2$ so that our master Lagrangian takes the final form
\begin{equation}
\mathcal{L}_{M}=\frac{\chi\theta}{2}\Pi_{\mu}\epsilon^{\mu\rho\sigma\delta}\partial_{\rho}B_{\sigma\delta}+\frac{m^{2}}{2}\Pi_{\mu}\Pi^{\mu}+\frac{\chi\theta}{2}\Lambda_{\mu\nu}\epsilon^{\mu\nu\rho\sigma}\partial_{\rho}A_{\sigma}-\frac{1}{4}\Lambda_{\mu\nu}\Lambda^{\mu\nu}-\frac{\chi\theta}{2}\varepsilon_{\mu\nu\alpha\beta}B^{\mu\nu}\partial^{\alpha}A^{\beta}.\label{Lmaster2}
\end{equation}

Accordingly, the Lagrangian density (\ref{Lmaster2}) describes both (\ref{SDmodel}) and (\ref{mkr}). This mechanism transforms models without gauge invariance into models with this symmetry by adding terms which does not appear on-shell. Note that the gauge invariance of $\mathcal{L}_{M}$ under $\delta A_{\mu} = \partial_{\mu}\lambda$ and $\delta B_{\mu\nu}=\partial_{\mu}\beta_{\nu}-\partial_{\nu}\beta_{\mu}$ with $\delta\Pi_{\mu}=\delta\Lambda_{\mu\nu} = 0$ is now evident, while it was a hidden symmetry in the self-dual formulation. With the master method, we were able to establish the relation of equivalence  when the coupling to other dynamical fields is considered and we have a simple formalism which account for the investigation of the theory at the quantum level.


\section{Duality mapping with a linear matter coupling\label{section3}}

The discussion on the duality developed in the previous section deals
only with free theories. However, it is fundamental to ensure that
this dual equivalence is also valid in the presence of external sources
coupled to the fields in $\mathcal{L}_{M}$. Here and throughout the
paper, we will assume only linear couplings with external fields,
whose associated currents are composed only of matter fields, represented
generically by $\psi$. The cases involving nonlinear couplings or
when the currents depend explicitly on the gauge or self-dual fields
are beyond our present scope.

Let us consider the master Lagrangian (\ref{Lmaster2}) added by dynamical matter
fields $\psi$ linearly coupled to the self-dual sector:
\begin{align}
\mathcal{L}_{M}^{(1)} & =\frac{\chi\theta}{2}\Pi_{\mu}\epsilon^{\mu\rho\sigma\delta}\partial_{\rho}B_{\sigma\delta}+\frac{m^{2}}{2}\Pi_{\mu}\Pi^{\mu}+\frac{\chi\theta}{2}\Lambda_{\mu\nu}\epsilon^{\mu\nu\rho\sigma}\partial_{\rho}A_{\sigma}\nonumber \\
 & -\frac{1}{4}\Lambda_{\mu\nu}\Lambda^{\mu\nu}-\frac{\chi\theta}{2}\varepsilon_{\mu\nu\alpha\beta}B^{\mu\nu}\partial^{\alpha}A^{\beta}+\Pi_{\mu}J^{\mu}+\Lambda_{\mu\nu}\mathcal{J}^{\mu\nu}+\mathcal{L}(\psi),\label{eq:LmasterS1}
\end{align}
where $\mathcal{L}(\psi)$ represents a generic Lagrangian density
responsible for the dynamics of the matter fields, with the corresponding
currents being denoted by $J_{\mu}$ and $\mathcal{J}_{\mu\nu}$.
Note that due to the lack of gauge symmetry in the self-dual sector,
the matter currents $J_{\mu}$ and $\mathcal{J}_{\mu\nu}$ are generally
not conserved. Also, to make our analysis as general as possible,
we will not assume any specific form to the matter sector for now.

First, we will remove the dependency on the gauge fields in Eq. (\ref{eq:LmasterS1}).
Varying the action $\int d^{4}x\mathcal{L}_{M}^{(1)}$ with respect
to the fields $A_{\mu}$ and $B_{\mu\nu}$, we obtain their corresponding
equations of motion whose solutions are given by
\begin{align}
A_{\mu} & =\Pi_{\mu}+\partial_{\mu}\phi,\\
B_{\mu\nu} & =\Lambda_{\mu\nu}+\partial_{\mu}\Sigma_{\nu}-\partial_{\nu}\Sigma_{\mu},
\end{align}
and substituting these solutions into Eq. (\ref{eq:LmasterS1}) we
find $\mathcal{L}_{M}^{(1)}=\mathcal{L}_{SD}^{(1)}$, with
\begin{align}
\mathcal{L}_{SD}^{(1)} & =\frac{m^{2}}{2}\Pi_{\mu}\Pi^{\mu}-\frac{1}{4}\Lambda_{\mu\nu}\Lambda^{\mu\nu}+\frac{\chi\theta}{2}\Pi_{\mu}\epsilon^{\mu\nu\alpha\beta}\partial_{\nu}\Lambda_{\alpha\beta}\nonumber \\
 & +\Pi_{\mu}J^{\mu}+\Lambda_{\mu\nu}\mathcal{J}^{\mu\nu}+\mathcal{L}(\psi).\label{eq:LSDSource1}
\end{align}
Then, $\mathcal{L}_{SD}^{(1)}$ is equivalent to the self-dual theory
(\ref{SDmodel}) linearly coupled to the matter, as expected. 

Next, we will eliminate the fields $\Pi_{\mu}$ and $\Lambda_{\mu\nu}$
from the master Lagrangian $\mathcal{L}_{M}^{(1)}$. The equations
of motion for these fields are
\begin{align}
\Pi_{\mu} & =-\frac{\chi\theta}{2m^{2}}\epsilon_{\mu\nu\alpha\beta}\partial^{\nu}B^{\alpha\beta}-\frac{1}{m^{2}}J_{\mu},\label{eq:PiJ}\\
\Lambda_{\mu\nu} & =\chi\theta\epsilon_{\mu\nu\alpha\beta}\partial^{\alpha}A^{\beta}+2\mathcal{J}_{\mu\nu}.\label{eq:LambdaJ}
\end{align}
Replacing Eqs. (\ref{eq:PiJ}) and (\ref{eq:LambdaJ}) into the master
Lagrangian then implies $\mathcal{L}_{M}^{(1)}=\mathcal{L}_{TM}^{(1)}$,
with
\begin{align}
\mathcal{L}_{TM}^{(1)} & =\frac{\theta^{2}}{12m^{2}}H_{\mu\nu\alpha}H^{\mu\nu\alpha}-\frac{\theta^{2}}{4}F_{\mu\nu}F^{\mu\nu}-\frac{\chi\theta}{2}\epsilon_{\mu\nu\alpha\beta}B^{\mu\nu}\partial^{\alpha}A^{\beta}\nonumber \\
 & -\frac{\chi\theta}{2m^{2}}B_{\mu\nu}\epsilon^{\mu\nu\alpha\beta}\partial_{\alpha}J_{\beta}-\frac{1}{2m^{2}}J_{\mu}J^{\mu}\nonumber \\
 & +\chi\theta A_{\mu}\epsilon^{\mu\nu\alpha\beta}\partial_{\nu}\mathcal{J}_{\alpha\beta}+\mathcal{J}_{\mu\nu}\mathcal{J}^{\mu\nu}+\mathcal{L}(\psi).\label{eq:LTMSource1}
\end{align}
From the above result, it is clear that the Lagrangian density $\mathcal{L}_{TM}^{(1)}$
represents the $TM_{B\wedge F}$ theory (\ref{mkr}) interacting with the matter through
``magnetic'' currents plus Thirring-like terms involving only the
matter fields. A similar Lagrangian density to the $\mathcal{L}_{TM}^{(1)}$
has appeared before in \cite{BF_dual}. However, the approach used
in \cite{BF_dual} was based on the gauge embedding method, different
from the one developed here. Also, one may verify that the equations
of motion for the fields $\Pi_{\mu}$ and $\Lambda_{\mu\nu}$ in the
$SD_{B\wedge F}$ model (\ref{eq:LSDSource1}) and for the gauge fields $A_{\mu}$
and $B_{\mu\nu}$ in the $TM_{B\wedge F}$ model (\ref{eq:LTMSource1}) can be
cast in the same form by means of the identification
\begin{align}
\Pi_{\mu} & \rightarrow\tilde{H}_{\mu}-\frac{1}{m^{2}}J_{\mu},\label{eq:PiJ2-1}\\
\Lambda_{\mu\nu} & \rightarrow\tilde{F}_{\mu\nu}+2\mathcal{J}_{\mu\nu}.\label{eq:LambdaJ2-1}
\end{align}

It is worth noting that the duality symmetry between $SD_{B\wedge F}/TM_{B\wedge F}$ theories
exchanges linear couplings $\Pi_{\mu}J^{\mu}$ and $\Lambda_{\mu\nu}\mathcal{J}^{\mu\nu}$,
involving currents not necessarily conserved in the self-dual sector
into derivative dual couplings $A_{\mu}\epsilon^{\mu\nu\alpha\beta}\partial_{\nu}\mathcal{J}_{\alpha\beta}$
and $B_{\mu\nu}\epsilon^{\mu\nu\alpha\beta}\partial_{\alpha}J_{\beta}$
in the gauge sector, whose associated currents are automatically conserved.
Moreover, self-interaction matter terms are naturally generated, which
will play a decisive role in ensuring the duality in the matter sector,
as we shall see in what follows.

\subsection{The matter sector}

Classically, the duality mapping established in Eqs. (\ref{eq:PiJ2-1}-\ref{eq:LambdaJ2-1})
ensures that the Lagrangian densities (\ref{eq:LSDSource1}) and (\ref{eq:LTMSource1})
are equivalent since the $SD_{B\wedge F}$ and $TM_{B\wedge F}$ fields obey the same equations
of motion in the presence of external sources. However, for this equivalence
between the models to be complete, it is also necessary to verify
what happens in the matter sector, when these sources are dynamics. 

To this end, we now consider the equation of motion for the matter
field $\psi$. First, let us focus our attention on the $SD_{B\wedge F}$ model
described by (\ref{eq:LSDSource1}), so 
\begin{equation}
\frac{\delta}{\delta\psi}\int d^{4}x\mathcal{L}_{SD}^{(1)}=0\Rightarrow\frac{\delta\mathcal{L}(\psi)}{\delta\psi}=-\Pi_{\mu}\frac{\delta J^{\mu}}{\delta\psi}-\Lambda_{\mu\nu}\frac{\delta\mathcal{J}^{\mu\nu}}{\delta\psi},\label{eq:Matter1}
\end{equation}
where $\frac{\delta\mathcal{L}(\psi)}{\delta\psi}$ is the Lagrangian
derivative.

On the other hand, the equations of motion for the fields $\Pi_{\mu}$
and $\Lambda_{\mu\nu}$ are:
\begin{eqnarray}
m^{2}\Pi^{\mu}+\frac{\chi\theta}{2}\epsilon^{\mu\nu\alpha\beta}\partial_{\nu}\Lambda_{\alpha\beta} & =-J^{\mu},\label{eq:PiJ2}\\
\frac{1}{2}\Lambda^{\mu\nu}-\frac{\chi\theta}{2}\epsilon^{\mu\nu\alpha\beta}\partial_{\alpha}\Pi_{\beta} & =\mathcal{J}^{\mu\nu},\label{eq:LambdaJ2}
\end{eqnarray}
and obey the constraints
\begin{align}
m^{2}\partial_{\mu}\Pi^{\mu} & =-\partial_{\mu}J^{\mu},\\
\partial_{\mu}\Lambda^{\mu\nu} & =2\partial_{\mu}\mathcal{J}^{\mu\nu}.
\end{align}
Inserting (\ref{eq:LambdaJ2}) into (\ref{eq:PiJ2}), we can eliminate
$\Lambda_{\mu\nu}$ in favor of $\Pi_{\mu}$ and obtain a second-order
differential equation as
\begin{equation}
\left(\theta^{2}\Box+m^{2}\right)\Pi^{\mu}=-J^{\mu}-\frac{\theta^{2}}{m^{2}}\partial^{\mu}\partial_{\nu}J^{\nu}-\chi\theta\epsilon^{\mu\nu\alpha\beta}\partial_{\nu}\mathcal{J}_{\alpha\beta},
\end{equation}
where we used the constraint $m^{2}\partial_{\mu}\Pi^{\mu}=-\partial_{\mu}J^{\mu}$.
Defining the wave-operator as $\hat{R}^{-1}=\Box+\frac{m^{2}}{\theta^{2}}$,
we can write
\begin{equation}
\Pi_{\mu}=-\frac{\hat{R}}{\theta^{2}}\left(J_{\mu}+\frac{\theta^{2}}{m^{2}}\partial_{\mu}\partial^{\nu}J_{\nu}+\chi\theta\epsilon_{\mu\nu\alpha\beta}\partial^{\nu}\mathcal{J}^{\alpha\beta}\right).\label{eq:PiJ3}
\end{equation}

A similar procedure for the field $\Lambda_{\mu\nu}$ results in 
\begin{equation}
\Lambda_{\mu\nu}=-\frac{\hat{R}}{\theta^{2}}\left[-2m^{2}\mathcal{J}_{\mu\nu}+2\theta^{2}\partial^{\alpha}\left(\partial_{\mu}\mathcal{J}_{\nu\alpha}-\partial_{\nu}\mathcal{J}_{\mu\alpha}\right)+\chi\theta\epsilon_{\mu\nu\alpha\beta}\partial^{\alpha}J^{\beta}\right].\label{eq:LambdaJ3}
\end{equation}

Replacing the solutions (\ref{eq:PiJ3}) and (\ref{eq:LambdaJ3})
back in the matter equation (\ref{eq:Matter1}), we come to the result
\begin{align}
\frac{\delta\mathcal{L}(\psi)}{\delta\psi} & =\frac{\hat{R}}{\theta^{2}}\left[J_{\mu}+\frac{\theta^{2}}{m^{2}}\partial_{\mu}\left(\partial^{\nu}J_{\nu}\right)+\chi\theta\epsilon_{\mu\nu\alpha\beta}\partial^{\nu}\mathcal{J}^{\alpha\beta}\right]\frac{\delta J^{\mu}}{\delta\psi}\nonumber \\
 & +\frac{\hat{R}}{\theta^{2}}\left[-2m^{2}\mathcal{J}_{\mu\nu}+2\theta^{2}\partial^{\alpha}\left(\partial_{\mu}\mathcal{J}_{\nu\alpha}-\partial_{\nu}\mathcal{J}_{\mu\alpha}\right)+\chi\theta\epsilon_{\mu\nu\alpha\beta}\partial^{\alpha}J^{\beta}\right]\frac{\delta\mathcal{J}^{\mu\nu}}{\delta\psi}.\label{eq:MatterSD}
\end{align}
This is a non-local differential equation, expressed only in terms
of the matter fields. 

Now, if we start from $\mathcal{L}_{TM}^{(1)}$, the equation of motion
for the matter field takes the form 
\begin{align}
\frac{\delta}{\delta\psi}\int d^{4}x\mathcal{L}_{TM}^{(1)} & =0\Rightarrow\frac{\delta\mathcal{L}(\psi)}{\delta\psi}=\left(\frac{1}{m^{2}}J_{\mu}-\hat{H}_{\mu}\right)\frac{\delta J^{\mu}}{\delta\psi}\nonumber \\
 & +\left(-2\mathcal{J}_{\mu\nu}-\hat{F}_{\mu\nu}\right)\frac{\delta\mathcal{J}^{\mu\nu}}{\delta\psi},\label{eq:Matter2}
\end{align}
where we have used the definitions (\ref{DualH}-\ref{DualF}) for the dual fields.

To eliminate the dual fields in (\ref{eq:Matter2}), we write the
equations of motion for $A_{\mu}$ and $B_{\mu\nu}$, obtained from
$\mathcal{L}_{TM}^{(1)}$, as 
\begin{align}
m^{2}\tilde{H}_{\mu}+\frac{\theta}{2\chi}\epsilon_{\mu\nu\alpha\beta}\partial^{\nu}\tilde{F}^{\alpha\beta} & =-\chi\theta\epsilon_{\mu\nu\alpha\beta}\partial^{\nu}\mathcal{J}^{\alpha\beta},\\
-\tilde{F}_{\mu\nu}+\frac{\theta}{\chi}\epsilon_{\mu\nu\alpha\beta}\partial^{\alpha}\tilde{H}^{\beta} & =\frac{\chi\theta}{m^{2}}\epsilon_{\mu\nu\alpha\beta}\partial^{\alpha}J^{\beta}.
\end{align}
These equations can be decoupled, and after some algebraic manipulations
we get the following results 
\begin{align}
\tilde{H}_{\mu} & =\frac{\hat{R}}{\theta^{2}}\left[\frac{\theta^{2}}{m^{2}}\left(\Box J_{\mu}-\partial_{\mu}\partial^{\nu}J_{\nu}\right)-\chi\theta\epsilon_{\mu\nu\alpha\beta}\partial^{\nu}\mathcal{J}^{\alpha\beta}\right],\\
\tilde{F}_{\mu\nu} & =-2R\Box\mathcal{J}_{\mu\nu}-\frac{\hat{R}}{\theta\text{\texttwosuperior}}\left[2\theta^{2}\partial^{\alpha}\left(\partial_{\mu}\mathcal{J}_{\nu\alpha}-\partial_{\nu}\mathcal{J}_{\mu\alpha}\right)+\chi\theta\epsilon_{\mu\nu\alpha\beta}\partial^{\alpha}J^{\beta}\right].
\end{align}
Substituting these solutions in Eq. (\ref{eq:Matter2}) we obtain 
\begin{align}
\frac{\delta\mathcal{L}(\psi)}{\delta\psi} & =\left[\frac{1}{m^{2}}\left(1-\hat{R}\Box\right)J_{\mu}+\frac{\hat{R}}{\theta^{2}}\left(\frac{\theta^{2}}{m^{2}}\partial_{\mu}\partial^{\nu}J_{\nu}+\chi\theta\epsilon_{\mu\nu\alpha\beta}\partial^{\nu}\mathcal{J}^{\alpha\beta}\right)\right]\frac{\delta J^{\mu}}{\delta\psi}\nonumber \\
 & +\left[2\left(\hat{R}\Box-1\right)\mathcal{J}_{\mu\nu}+\frac{\hat{R}}{\theta^{2}}\left(2\theta^{2}\partial^{\alpha}\left(\partial_{\mu}\mathcal{J}_{\nu\alpha}-\partial_{\nu}\mathcal{J}_{\mu\alpha}\right)+\chi\theta\epsilon_{\mu\nu\alpha\beta}\partial^{\alpha}J^{\beta}\right)\right]\frac{\delta\mathcal{J}^{\mu\nu}}{\delta\psi}.
\end{align}
Using the definition $\hat{R}^{-1}=\Box+\frac{m^{2}}{\theta^{2}}$,
we can write $\Box=R^{-1}-\frac{m^{2}}{\theta^{2}}$ which implies 
\begin{align}
\frac{\delta\mathcal{L}(\psi)}{\delta\psi} & =\frac{\hat{R}}{\theta^{2}}\left[J_{\mu}+\frac{\theta^{2}}{m^{2}}\partial_{\mu}\left(\partial^{\nu}J_{\nu}\right)+\chi\theta\epsilon_{\mu\nu\alpha\beta}\partial^{\nu}\mathcal{J}^{\alpha\beta}\right]\frac{\delta J^{\mu}}{\delta\psi}\nonumber \\
 & +\frac{\hat{R}}{\theta^{2}}\left[-2m^{2}\mathcal{J}_{\mu\nu}+2\theta^{2}\partial^{\alpha}\left(\partial_{\mu}\mathcal{J}_{\nu\alpha}-\partial_{\nu}\mathcal{J}_{\mu\alpha}\right)+\chi\theta\epsilon_{\mu\nu\alpha\beta}\partial^{\alpha}J^{\beta}\right]\frac{\delta\mathcal{J}^{\mu\nu}}{\delta\psi}.\label{eq:MatterTM}
\end{align}

By comparing Eqs. (\ref{eq:MatterSD}) and (\ref{eq:MatterTM}), we
conclude that the matter sectors of the two models give rise to the
same equations of motion. Thus, we have shown that the Lagrangians
$\mathcal{L}_{SD}^{(1)}$ and $\mathcal{L}_{TM}^{(1)}$ are equivalent
and have established the classical duality between the $SD_{B\wedge F}$ and $TM_{B\wedge F}$
theories when couplings with dynamical matter fields are considered. 

In order to liken our results with the literature, let us consider,
as a particular case, a fermionic matter field minimally coupled to
the self-dual field $\Pi_{\mu}$. Assuming the following identifications:
\begin{equation}
\mathcal{L}(\psi)\rightarrow\mathcal{L}_{Dirac}=\bar{\psi}(i\gamma^{\mu}\partial_{\mu}-M)\psi,
\end{equation}
where $M$ is the Dirac field mass, and the fermionic currents are 
\begin{align}
J_{\mu} & \rightarrow-eJ_{\mu}=-e\bar{\psi}\gamma_{\mu}\psi,\\
\mathcal{J}_{\mu\nu} & \rightarrow0,
\end{align}
with $e$ being a dimensionless coupling constant. The equation of
motion for $\psi$ (\ref{eq:MatterTM}) takes the simple form 
\begin{equation}
(i\gamma^{\mu}\partial_{\mu}-M)\psi=\frac{e^{2}}{\theta^{2}}\hat{R}J_{\mu}\gamma^{\mu}\psi,
\end{equation}
which agrees with the result obtained in \cite{BF_dual}.

\section{The duality at the quantum level\label{section4}}

Once we proved the duality between $SD_{B \wedge F}$ and $TM_{B \wedge F}$ models at the level
of equations of motion, we now check whether this duality is preserved
at the quantum level. For this purpose, we adopt the path-integral
framework and define the master generating functional as
\begin{equation}
Z(\psi)=\mathcal{N}\int\mathcal{D}A^{\mu}\mathcal{D}B^{\mu\nu}\mathcal{D}\Pi^{\mu}\mathcal{D}\Lambda^{\mu\nu}\exp\left\{ i\int d^{4}x\left[\mathcal{L}_{M}+J_{\mu}\Pi^{\mu}+\mathcal{J}_{\mu\nu}\Lambda^{\mu\nu}+\mathcal{L}(\psi)\right]\right\} ,\label{eq:Z1}
\end{equation}
where $\mathcal{N}$ is a overall normalization constant. Our aim
is to evaluate the effective Lagrangian resulting from the integration
over the fields. Firstly, let us integrate out the contribution of
the $SD_{B\wedge F}$ fields.

After the shifts, $\Pi_{\mu}\rightarrow\Pi_{\mu}+\tilde{H}_{\mu}-\frac{1}{m^{2}}J_{\mu}$
and $\Lambda_{\mu\nu}\rightarrow\Lambda_{\mu\nu}+\tilde{F}_{\mu\nu}+2\mathcal{J}_{\mu\nu}$, 
we perform the functional integration in Eq. (\ref{eq:Z1}) over
the fields $\Pi_{\mu}$ and $\Lambda_{\mu\nu}$, thereby producing 
\begin{equation}
Z(\psi)=\mathcal{N}\int\mathcal{D}A^{\mu}\mathcal{D}B^{\mu\nu}\exp\left[i\int d^{4}x\mathcal{L}_{eff}^{(1)}(A,B,\psi)\right],\label{eq:Z2}
\end{equation}
where
\begin{align}
\mathcal{L}_{eff}^{(1)}(A,B,\psi) & =\frac{\theta^{2}}{12m^{2}}H_{\mu\nu\alpha}H^{\mu\nu\alpha}-\frac{\theta^{2}}{4}F_{\mu\nu}F^{\mu\nu}-\frac{\chi\theta}{2}\epsilon_{\mu\nu\alpha\beta}B^{\mu\nu}\partial^{\alpha}A^{\beta}\nonumber \\
 & -\frac{\chi\theta}{2m^{2}}B_{\mu\nu}\epsilon^{\mu\nu\alpha\beta}\partial_{\alpha}J_{\beta}-\frac{1}{2m^{2}}J_{\mu}J^{\mu}\nonumber \\
 & +\chi\theta A_{\mu}\epsilon^{\mu\nu\alpha\beta}\partial_{\nu}\mathcal{J}_{\alpha\beta}+\mathcal{J}_{\mu\nu}\mathcal{J}^{\mu\nu}+\mathcal{L}(\psi),\label{eq:LMaster1-2}
\end{align}
is the same Lagrangian density found in Eq. (\ref{eq:LTMSource1}). 

To integrate over the fields configurations $A_{\mu}$ and $B_{\mu\nu}$,
let us first note that the master Lagrangian $\mathcal{L}_{M}$ can
be rewritten, up to surface terms, as
\begin{equation}
\mathcal{L}_{M}=\frac{\chi\theta}{2}\epsilon^{\mu\nu\alpha\beta}\left(\Lambda_{\mu\nu}-B_{\mu\nu}\right)\partial_{\alpha}\left(A_{\beta}-\Pi_{\beta}\right)+\mathcal{L}_{SD}.\label{eq:LMaster2-1}
\end{equation}
 In this way, we can make a shift in the gauge fields through $B_{\mu\nu}\rightarrow B_{\mu\nu}+\Lambda_{\mu\nu}$
and $A_{\beta}\rightarrow A_{\beta}+\Pi_{\beta}$, which allows us
to rewrite the generating function (\ref{eq:Z1}) as, 
\begin{equation}
Z(\psi)=\mathcal{N}\int\mathcal{D}A^{\mu}\mathcal{D}B^{\mu\nu}\mathcal{D}\Pi^{\mu}\mathcal{D}\Lambda^{\mu\nu}\exp\left\{ i\int d^{4}x\left[-\frac{\chi\theta}{2}\epsilon^{\mu\nu\alpha\beta}B_{\mu\nu}\partial_{\alpha}A_{\beta}+\mathcal{L}_{SD}+J_{\mu}\Pi^{\mu}+\mathcal{J}_{\mu\nu}\Lambda^{\mu\nu}+\mathcal{L}(\psi)\right]\right\} ,
\end{equation}
such that the $A_{\mu}$ and $B_{\mu\nu}$ fields decouple. Then, performing the function integration yields to the following generating functional
\begin{equation}
Z(\psi)=\mathcal{N}\int\mathcal{D}\Pi^{\mu}\mathcal{D}\Lambda^{\mu\nu}\exp\left[i\int d^{4}x\mathcal{L}_{eff}^{(2)}(\Pi,\Lambda,\psi)\right],\label{eq:Z3}
\end{equation}
with
\begin{align}
\mathcal{L}_{eff}^{(2)}(\Pi,\Lambda,\psi) & =\frac{m^{2}}{2}\Pi_{\mu}\Pi^{\mu}-\frac{1}{4}\Lambda_{\mu\nu}\Lambda^{\mu\nu}+\frac{\chi\theta}{2}\Pi_{\mu}\epsilon^{\mu\nu\alpha\beta}\partial_{\nu}\Lambda_{\alpha\beta}\nonumber \\
 & +\Pi_{\mu}J^{\mu}+\Lambda_{\mu\nu}\mathcal{J}^{\mu\nu}+\mathcal{L}(\psi),\label{eq:LMaster2-2}
\end{align}
corresponding to the same Lagrangian density (\ref{eq:LSDSource1})
previously obtained. It is worth highlighting the physical implications
contained in (\ref{eq:LMaster2-1}). We clearly see that the master
Lagrangian $\mathcal{L}_{M}$ obtained in (\ref{Lmaster2}) is equivalent to self-dual
Lagrangian $\mathcal{L}_{SD}$ added by a purely topological $B\wedge F$
term, which makes evident the role of the master Lagrangian on the duality symmetry. 

The implications of the above results at the quantum level can be explored by considering
the functional derivatives of (\ref{eq:Z2}) and (\ref{eq:Z3})
with respect to the sources. Setting $J_{\mu}=\mathcal{J_{\mu\nu}}=0$,
we can establish the following identities to the correlation functions
\begin{align}
\left\langle \Pi_{\mu_{1}}(x_{1})\cdots\Pi_{\mu_{N}}(x_{N})\right\rangle _{SD} & =\left\langle \tilde{H}_{\mu_{1}}\left[B(x_{1})\right]\cdots\tilde{H}_{\mu_{N}}\left[B(x_{N})\right]\right\rangle _{TM}+\mbox{contact terms},\\
\left\langle \Lambda_{\mu_{1}\nu_{1}}(x_{1})\cdots\Lambda_{\mu_{N}\nu_{N}}(x_{N})\right\rangle _{SD} & =\left\langle \tilde{F}_{\mu_{1}\nu_{1}}\left[A(x_{1})\right]\cdots\tilde{F}_{\mu_{N}\nu_{N}}\left[A(x_{N})\right]\right\rangle _{TM}+\mbox{contact terms}.
\end{align}
These relations show that the classical dual map (\ref{eq:PiJ2-1}-\ref{eq:LambdaJ2-1})
is satisfied by all quantum correlation functions of those fields,
up to contact terms. 

Finally, we now complete the proof of quantum duality between the
$SD_{B\wedge F}/TM_{B\wedge F}$ models by performing the path integration over $A_{\mu}$
and $B_{\mu\nu}$ gauge fields in Eq. (\ref{eq:Z2}), and over $\Pi_{\mu}$
and $\Lambda_{\mu\nu}$ self-dual fields in Eq. (\ref{eq:Z3}). For
this goal, it is convenient to organize the effective Lagrangians
(\ref{eq:LMaster1-2}) and (\ref{eq:LMaster2-2}) in a matrix-form
according to the
\begin{equation}
\mathcal{L}=\frac{1}{2}\boldsymbol{X}^{T}\mathcal{\hat{O}}\boldsymbol{X}+\mathbf{X}^{T}\boldsymbol{J},
\end{equation}
where the wave operator, $\mathcal{\hat{O}}$, form a $2\times2$
matrix, $\boldsymbol{X}$, and $\boldsymbol{J}$ represent vector-tensor
duplet of type
\begin{equation}
\boldsymbol{X}=\begin{pmatrix}A_{\mu}\\
B_{\mu\nu}
\end{pmatrix}.
\end{equation}

To accomplish the functional integration, we use the Gaussian path
integral formula over a bosonic field $\boldsymbol{X}$,
\begin{equation}
\int\mathcal{D}\boldsymbol{X}\exp\left[i\int d^{4}x\left(\frac{1}{2}\boldsymbol{X}^{T}\mathcal{\hat{O}}\boldsymbol{X}+\mathbf{X}^{T}\boldsymbol{J}\right)\right]=\left[\mbox{Det}\left(-i\mathcal{\hat{O}}\right)\right]^{-\frac{1}{2}}\times\exp\left[-i\int d^{4}x\frac{1}{2}\boldsymbol{J}^{T}\hat{\mathcal{O}}^{-1}\boldsymbol{J}\right].\label{eq:IntGaussian}
\end{equation}

In our case, the determinant $\mbox{Det}\left(-i\mathcal{\hat{O}}\right)$
is field-independent and can be absorbed by the normalization constant.
The calculation of propagators $\hat{\mathcal{O}}^{-1}$ is rather
lengthy, and the details are in Appendix \ref{Appendix}. Here we just write the results
\begin{equation}
\left(\mathcal{\hat{O}}_{SD}^{-1}\right)^{\mu,\alpha\beta;\nu,\lambda\sigma}=\left(\begin{array}{cc}
\frac{1}{\theta^{2}\Box+m^{2}}\Theta^{\mu\nu}+\frac{1}{m^{2}}\omega^{\mu\nu} & \frac{2}{\theta^{2}\Box+m^{2}}S^{\mu\lambda\sigma}\\
-\frac{2}{\theta^{2}\Box+m^{2}}S^{\alpha\beta\nu} & -\frac{2m^{2}}{\theta^{2}\Box+m^{2}}\left(P^{(1)}\right){}^{\alpha\beta,\lambda\sigma}-2\left(P^{(2)}\right){}^{\alpha\beta,\lambda\sigma}
\end{array}\right),
\end{equation}
and
\begin{equation}
\left(\hat{\mathcal{O}}_{TM}^{-1}\right)^{\mu,\alpha\beta;\nu,\lambda\sigma}=\left(\begin{array}{cc}
\frac{1}{\theta^{2}\Box+m^{2}}\Theta^{\mu\nu}+\frac{\lambda}{\Box}\omega^{\mu\nu} & -\frac{2m^{2}}{\theta^{2}\Box\left(\theta^{2}\Box+m^{2}\right)}S^{\mu\lambda\sigma}\\
\frac{2m^{2}}{\theta^{2}\Box\left(\theta^{2}\Box+m^{2}\right)}S^{\alpha\beta\nu} & -\frac{2m^{2}}{\theta^{2}\Box+m^{2}}\left(P^{(1)}\right){}^{\alpha\beta,\lambda\sigma}-\frac{2\xi}{\Box}\left(P^{(2)}\right){}^{\alpha\beta,\lambda\sigma}
\end{array}\right),
\end{equation}
where $\Theta_{\mu\nu}$, $\omega_{\mu\nu}$, $S_{\mu\nu\alpha}$,
$P_{\mu\nu,\alpha\beta}^{(1)}$ and $P_{\mu\nu,\alpha\beta}^{(2)}$
are projection operators whose definitions and closed algebras are
shown in Appendix \ref{Appendix}. Also, $\lambda$ and $\xi$ are convenient gauge
fixing parameters. Note that the physical poles of the two propagators
are equal, i.e., $\theta^{2}\Box+m^{2}=0$, and confirm that the particle
spectrum of both theories are equivalent, so that we may consider
the self-dual theory equivalent to $TM_{B\wedge F}$ theory with the fixed gauge.

The above propagators, together with formula (\ref{eq:IntGaussian}),
enable us to perform the functional integration in (\ref{eq:Z2})
and (\ref{eq:Z3}). After completing all tensorial contractions, we
obtain the same effective Lagrangian for the matter field
\begin{align}
\mathcal{L}_{eff}^{(3)}(\psi) & =\mathcal{L}(\psi)\nonumber \\
 & +\frac{1}{2}\begin{pmatrix}J_{\mu} & \mathcal{J_{\alpha\beta}}\end{pmatrix}\begin{pmatrix}-\frac{1}{\theta^{2}\Box+m^{2}}\eta^{\mu\nu}-\frac{\theta^{2}}{m^{2}}\frac{1}{\theta^{2}\Box+m^{2}}\partial^{\mu}\partial^{\nu} & -\frac{\chi\theta}{\theta^{2}\Box+m^{2}}\epsilon^{\mu\lambda\sigma\delta}\partial_{\delta}\\
\frac{\chi\theta}{\theta^{2}\Box+m^{2}}\epsilon^{\alpha\beta\nu\delta}\partial_{\delta} & \frac{2}{\theta^{2}\Box+m^{2}}\left(\theta^{2}\Box P^{(2)}+m^{2}\mathcal{I}\right){}^{\alpha\beta,\lambda\sigma}
\end{pmatrix}\begin{pmatrix}J_{\nu}\\
\mathcal{J}_{\lambda\sigma}
\end{pmatrix}.\label{eq:Leff3}
\end{align}

It is easy to verify that the equation of motion for the matter field
obtained from $\mathcal{L}_{eff}^{(3)}$ (\ref{eq:Leff3}) is precisely
that found in the previous section (see Eqs. (\ref{eq:MatterSD})
or (\ref{eq:MatterTM})). Thus, we prove the quantum equivalence between
the matter sector of the $SD_{B\wedge F}/TM_{B\wedge F}$ models. It is worth mentioning that
the dynamics of the matter fields is preserved in the functional integration
in (\ref{eq:Z2}) only if the Thirring-like interactions are added
to the diagonal elements of $\hat{\mathcal{O}}_{TM}^{-1}$ matrix.
Besides, the gauge-dependent parts involving the gauge fixing parameters
are canceled, as it should be.

\section{Conclusion\label{conclusion}}

In this work, we revisited the 
duality between the
self-dual and topologically massive models involving the $B\wedge F$
term in $3+1$ spacetime dimensions. The study of this duality when
couplings with fermionic matter are included was first carried out
in \cite{BF_dual}, through the gauge embedding formalism. Here, we
considered another approach, namely the master action method, whereby
we obtained a fundamental Lagrangian density that interpolates
between the two models and provides direct proof of dual equivalence
at both the classical and quantum level. The master action enabled us to relate the equations of motion of these models
via a dual map among fields and currents of both theories, which ensures
that they are equivalent at the classical level. In addition, we demonstrated the duality at
quantum through the path-integral framework.
We defined a master generating functional wherein the integration
over the different fields provided effective Lagrangians that
are the same as those obtained classically. Moreover, after a last
functional integration over the bosonic fields, we obtained an effective
non-local Lagrangian for the matter fields, which proves the equivalence
between the matter sectors of the analyzed models.

We assumed that the external currents are linearly coupled
with the self-dual fields and are constituted exclusively of the matter
fields. We show that these interactions induce ``magnetic''
couplings involving the gauge fields, in addition to current-current
Thirring-like interactions. These types of couplings are, in general,
non-renormalizable by direct power counting \cite{malacarne,Ferrari1}. However, as in $2+1$ dimensional case involving the Maxwell-Chern-Simons model, we may expect which this weakness can
be overcome by a $\nicefrac{1}{N}$ perturbative expansion when the
matter field is an $N$-component fermionic field, such that the theory
becomes renormalizable. An explicit verification
of this issue, as well as a possible extension of our results to the
supersymmetric case \cite{Ferrari2, Almeida}, are themes for forthcoming
works.

\section*{Acknowledgments}
\hspace{0.5cm}
The authors thank the Funda\c{c}\~{a}o Cearense de Apoio ao Desenvolvimento
Cient\'{i}fico e Tecnol\'{o}gico (FUNCAP), the Coordena\c{c}\~{a}o de Aperfei\c{c}oamento de Pessoal de N\'{i}vel Superior (CAPES), and the Conselho Nacional de Desenvolvimento Cient\'{i}fico e Tecnol\'{o}gico (CNPq), Grants no 312356/2017-0 (JEGS), no 305678/2015-9 (RVM) and no 308638/2015-8 (CASA) for financial support.

\appendix

\section{Feynman propagator for the $TM_{B\wedge F}$ theory\label{Appendix}}

Consider the topologically massive $B\wedge F$ model defined as 
\begin{equation}
S_{TM}=\int d^{4}x\left[-\frac{\theta^{2}}{4}F^{\mu\nu}F_{\mu\nu}+\frac{\theta^{2}}{12m^{2}}H^{\mu\nu\alpha}H_{\mu\nu\alpha}-\frac{\chi\theta}{4}\epsilon_{\mu\nu\alpha\beta}B^{\mu\nu}F^{\alpha\beta}\right],\label{A1}
\end{equation}
where the first two terms represent a gauge-invariant Maxwell-Kalb-Ramond
theory, while the last is a topological $B\wedge F$ term. The calculation
of the Feynman propagator for the theory (\ref{A1}) can be performed
as follows.

First, let us rewrite the integrand in Eq. (\ref{A1}) on the matrix
form
\begin{equation}
\mathcal{L}_{TM}=\frac{1}{2}\boldsymbol{X}^{T}\mathcal{\hat{O}}_{TM}\boldsymbol{X},\label{eq:A2}
\end{equation}
with the wave operator, $\mathcal{\hat{O}}_{TM}$, being a $2\times2$
matrix, and $\boldsymbol{X}$ represents a column vector-tensor as
\begin{equation}
\boldsymbol{X}=\begin{pmatrix}A_{\mu}\\
B_{\mu\nu}
\end{pmatrix}.
\end{equation}

Adding convenient gauge-fixing terms in (\ref{eq:A2}), namely, $-\frac{1}{2\lambda}\left(\partial_{\mu}A^{\mu}\right)^{2}$
and $\frac{1}{2\xi}\left(\partial_{\mu}B^{\mu\nu}\right)^{2}$, we
can explicitly write the operator $\mathcal{\hat{O}}_{TM+gf}$, in
the form, 
\begin{equation}
\mathcal{\hat{O}}_{TM+gf}^{\mu,\alpha\beta;\nu,\lambda\sigma}=\left(\begin{array}{cc}
\theta^{2}\Box\Theta^{\mu\nu}+\frac{\Box}{\lambda}\omega^{\mu\nu} & -S^{\mu\lambda\sigma}\\
S^{\alpha\beta\nu} & -\frac{\theta^{2}\Box}{2m^{2}}\left(P^{(1)}\right){}^{\alpha\beta,\lambda\sigma}-\frac{\Box}{2\xi}\left(P^{(2)}\right){}^{\alpha\beta,\lambda\sigma}
\end{array}\right),
\end{equation}
where we have introduced the set of spin-projection operators as
\begin{align}
\Theta_{\mu\nu} & =\eta_{\mu\nu}-\omega_{\mu\nu},\ \ \ \ \omega_{\mu\nu}=\frac{\partial_{\mu}\partial_{\nu}}{\Box},\\
S_{\mu\nu\alpha} & =\frac{\chi\theta}{2}\epsilon_{\mu\nu\alpha\beta}\partial^{\beta},\\
P_{\mu\nu,\alpha\beta}^{(1)} & =\frac{1}{2}\left(\Theta_{\mu\alpha}\Theta_{\nu\beta}-\Theta_{\mu\beta}\Theta_{\nu\alpha}\right),\\
P_{\mu\nu,\alpha\beta}^{(2)} & =\frac{1}{2}\left(\Theta_{\mu\alpha}\omega_{\nu\beta}-\Theta_{\mu\beta}\omega_{\nu\alpha}+\Theta_{\nu\beta}\omega_{\mu\alpha}-\Theta_{\nu\alpha}\omega_{\mu\beta}\right),
\end{align}
with $\Box\equiv\partial_{\mu}\partial^{\mu}$, and $\eta_{\mu\nu}$
is the Minkowski metric with signature $\left(+,-,-,-\right)$. Note
that $P^{(1)}$ and $P^{(2)}$ satisfy the tensorial completeness
relation:
\begin{equation}
\left(P^{(1)}+P^{(1)}\right)_{\mu\nu,\alpha\beta}=\frac{1}{2}\left(\eta_{\mu\alpha}\eta_{\nu\beta}-\eta_{\mu\beta}\eta_{\nu\alpha}\right)\equiv\mathcal{I}_{\mu\nu,\alpha\beta}.
\end{equation}

The products between the operators defined above satisfy a closed
algebra and are summarized in Tables \ref{table1}, \ref{table2}.

\begin{table}[ptb]
\begin{tabular}{|c|c|c|}
\hline 
 & $\Theta_{\ \nu}^{\alpha}$ & $\omega_{\ \nu}^{\nu}$\tabularnewline
\hline 
\hline 
$\Theta_{\mu\alpha}$ & $\Theta_{\mu\nu}$ & $0$\tabularnewline
\hline 
$\omega_{\mu\alpha}$ & $0$ & $\omega_{\mu\nu}$\tabularnewline
\hline 
\end{tabular}
\ \ 
\begin{tabular}{|c|c|c|}
\hline 
 & $\left(P^{(1)}\right)_{\ \ \alpha\beta}^{\rho\sigma}$ & $\left(P^{(2)}\right)_{\ \ \alpha\beta}^{\rho\sigma}$\tabularnewline
\hline 
\hline 
$P_{\mu\nu\rho\sigma}^{(1)}$ & $P_{\mu\nu\alpha\beta}^{(1)}$ & $0$\tabularnewline
\hline 
$P_{\mu\nu\rho\sigma}^{(2)}$ & $0$ & $P_{\mu\nu\alpha\beta}^{(2)}$\tabularnewline
\hline 
\end{tabular}
\caption{Algebra of the spin-projection operators.}%
\label{table1}%
\end{table}

\begin{table}[ptb]
\begin{tabular}{|c|c|c|c|}
\hline 
 & $S_{\alpha\beta\nu}$ & $\Theta_{\beta\sigma}$ & $\omega_{\beta\sigma}$\tabularnewline
\hline 
\hline 
$S^{\mu\alpha\beta}$ & $-\frac{\theta^{2}}{2}\Box\Theta_{\ \nu}^{\mu}$ & $S_{\ \ \  \sigma}^{\mu\alpha}$ & $0$\tabularnewline
\hline 
\end{tabular}
\ \ 
\begin{tabular}{|c|c|c|c|}
\hline 
 & $S_{\ \alpha\beta}^{\lambda}$ & $\left(P^{(1)}\right)_{\ \ \rho\sigma}^{\nu\lambda}$ & $\left(P^{(2)}\right)_{\ \ \rho\sigma}^{\nu\lambda}$\tabularnewline
\hline 
\hline 
$S_{\mu\nu\lambda}$ & $-\frac{\theta^{2}}{2}\Box P_{\mu\nu\alpha\beta}^{(1)}$ & $S_{\mu\rho\sigma}$ & $0$\tabularnewline
\hline 
\end{tabular}
\caption{Algebra of the spin-projection operators.}%
\label{table2}%
\end{table}

The Feynman propagator is defined as $\mathcal{\hat{O}}^{-1}_{TM+gf}$.
In order to invert the wave operator, we will write it and its inverse
generically by:

\begin{equation}
\mathbf{O}=\left(\begin{array}{cc}
A & B\\
C & D
\end{array}\right),\qquad\text{and}\qquad\mathbf{O}^{-1}=\left(\begin{array}{cc}
\mathbb{A} & \mathbb{B}\\
\mathbb{C} & \mathbb{D}
\end{array}\right),
\end{equation}
which fulfills the relation $\mathbf{O}\mathbf{O}^{-1}=\mathbf{I}$,
where the general identity matrix $\mathbf{I}$ is defined by:
\begin{equation}
\mathbf{I}=\left(\begin{array}{cc}
\textit{I} & 0\\
0 & \mathbb{\mathcal{I}}
\end{array}\right),
\end{equation}
with $\textit{I}$ and $\mathcal{I}$ are the identities to the projectors
$(\theta^{\mu\nu},$ $\omega^{\mu\nu})$, and $(P^{(1)}$, $P^{(2)})$,
respectively. From these preliminary definitions, we obtain a system of four equations, whose solutions
can be written as 
we get
\begin{equation}
\left\lbrace \begin{array}{l}
A\mathbb{A}+B\mathbb{C}=\textit{I}\\
A\mathbb{B}+B\mathbb{D}=0\\
C\mathbb{A}+D\mathbb{C}=0\\
C\mathbb{B}+D\mathbb{D}=\mathcal{I}
\end{array}\right.\Rightarrow\left\lbrace \begin{array}{l}
\mathbb{A}=(A-BD^{-1}C)^{-1}\\
\mathbb{B}=-A^{-1}B\mathbb{D}\\
\mathbb{C}=-D^{-1}C\mathbb{A}\\
\mathbb{D}=(D-CA^{-1}B)^{-1}
\end{array}\right.
\end{equation}

After some algebraic manipulations with the set of the operators presented
above, the $TM_{B\wedge F}$ gauge propagator is properly written
as 

\begin{equation}
\left(\hat{\mathcal{O}}_{TM}^{-1}\right)^{\mu,\alpha\beta;\nu,\lambda\sigma}=\left(\begin{array}{cc}
\frac{1}{\theta^{2}\Box+m^{2}}\Theta^{\mu\nu}+\frac{\lambda}{\Box}\omega^{\mu\nu} & -\frac{2m^{2}}{\theta^{2}\Box\left(\theta^{2}\Box+m^{2}\right)}S^{\mu\lambda\sigma}\\
\frac{2m^{2}}{\theta^{2}\Box\left(\theta^{2}\Box+m^{2}\right)}S^{\alpha\beta\nu} & -\frac{2m^{2}}{\theta^{2}\Box+m^{2}}\left(P^{(1)}\right){}^{\alpha\beta,\lambda\sigma}-\frac{2\xi}{\Box}\left(P^{(2)}\right){}^{\alpha\beta,\lambda\sigma}
\end{array}\right).
\end{equation}


\end{document}